\documentclass[review]{elsarticle}

\usepackage{lineno,hyperref}
\modulolinenumbers[5]

\journal{Journal of \LaTeX\ Templates}

\begin{document}

\begin{frontmatter}

\title{Propagation dynamics of vortices in Helico-Conical optical beams}

\author{Nestor Bareza Jr.$^*$ and Nathaniel Hermosa}
\address{$^1$National Instittute of Physics, University of the Philippines Diliman, 1101 Philippines}
\address{$^*$Corresponding author: nbareza@nip.upd.edu.ph}
\fntext[myfootnote]{1101 University of the Philippines Diliman, Quezon City, Philippines.}

%
%

\begin{abstract}
We present the dynamics of optical vortices (OVs) that came from the propagation of helico-conical optical beam. This dynamics is investigated numerically by tracking the OVs at several distances using rigorous scalar diffraction theory. To ensure that our numerical calculations are correct, we compare the intensity profiles and their corresponding interferograms taken at different propagation distances between simulations and experiments. We observe that the peripheral isopolar vortices transport radially inward, toward the optical axis along the transverse spatial space as the beam propagates. When the beam has a central vortex, these vortices have significant induced angular rates of motion about the optical axis. These propagation dynamics of vortices influence the internal energy flow and the wave profile reconstruction of the beam, which can be important when deciding their applications.
\end{abstract}

\begin{keyword}
\texttt{Optical Vortices, Wave propagation, Laser beam shaping}

\end{keyword}

\end{frontmatter}


\section{Introduction}
An optical vortex (OV), first described in the seminal work of Nye and Berry, is a region of singularity in which the wave amplitude vanishes and the phase is indeterminate \cite{nye1974dislocations}. Wavefronts that contain OV include Laguerre-Gaussian beams \cite{allen1992orbital} and higher order Bessel beams \cite{volke2002orbital}, as well as their fractional counterparts \cite{tao2004experimental, gotte2008light}. An underlying interest on these kinds of beams emerges since OV carries orbital angular momentum. These beams also have wide range of potential applications such as in information encoding, free-space information transfer \cite{gibson2004free}, optical trapping \cite{cojoc2005laser} and micromanipulation \cite{lee2004optical}.

The study of the dynamics of OV may offer substantial explanations to other related physical phenomena such as drift events, gyration and hydrodynamics \cite{roux1995dynamical, vaupel1996hydrodynamic}. Thus, there is a demand to understand the dynamical behavior of OV. OV has fluidlike motion as the beam propagates which is explained by the potential theory \cite{rozas1997experimental}. For multiple OVs in a single beam, same-charge (isopolar) vortices are found to gyrate together while opposite-charge (bipolar) vortices tend to drift away from the direction of beam axis \cite{roux1995dynamical}. Investigating dynamical behavior of three or more OVs in a single beam has not been fully explored. Although Roux derived mathematical expressions for interaction of two OVs \cite{roux1995dynamical}, several complications arise in providing a preliminary model for interacting three or more OVs. This kind of system, which cannot be described by mutual interactions such as 2-body mechanics problem and 2-charge electrodynamics problem, requires statistical approach \cite{ahmad2002statistical, sah1958electron}. 

Helico-conical optical beam (HCOB) is a special type of beam which can produce a string of multiple OVs when propagated. The HCOB is an interesting beam in that it acquires OVs as it propagates because of its peculiar phase: it has inseparable azimuthal $\theta$ and radial $\rho$ phases. HCOB phase $\phi(\rho,\theta)$ is expressed as:
\begin{equation}
\phi(\rho,\theta) = l\theta(K-\frac{\rho}{\rho_o})
\label{eqn:HCOBphase}
\end{equation}
where $l$  is the topological charge, $\rho_o$ normalizes $\rho$, and $K$ can take the value of either 0 or 1. HCOB has been observed to generate a string of $l$ isopolar peripheral OVs and has an oppositely $l$-charged central vortex for $K$ equal to 1 \cite{hermosa2007phase}. These same polarity OVs in the radial opening are interacting which cause angular and radial phase variations \cite{singh2013vortices}. Their interactions cause the rotations of OVs as the beam propagates which affects the intensity profile. In this paper, we report the first detailed study of dynamical behavior of a string of multiple OVs in HCOB as the beam propagates. We presented the interaction of peripheral OVs with and without the presence of central OV. This is based on the radial and angular motions of OVs in beam profiles that are obtained at fine-scaled propagation distances.

Several approaches in OV detection have been demonstrated numerically and experimentally, in which some are discussed in the following. Roux performed a numerical closed line integral to the gradient of the phase profile in locating the OV \cite{roux1995dynamical}. A closed line integral to a continuous gradient results to a zero value. Hence, an integral which results to a non-zero value implies that it is evaluated at discontinuous region, which indicates that inside the contour is an OV. Maallo and Almoro demonstrated detection of single OV by the axial behavior in retrieved phase maps \cite{maallo2011numerical}. They based their detection on the behavior of OV which has rapid phase variation in the transverse space and has phase invariance along axial direction. Murphy et al. experimentally located OVs using a Shack-Hartmann wavefront sensor \cite{murphy2010experimental}. The slopes in the acquired wavefront surface are used in an algorithm to display branch point potential map in which peaks and valleys display both the locations and polarities of the OVs.  Ricci et al. used multi-pinhole interferometry to demonstrate the vortex-splitting phenomenon that is able to detect OV when 2D scanning is performed \cite{ricci2012instability}. In our study, it is necessary that multiple OVs can be detected simultaneously even at relatively far propagation distance. Also, intensity profiles obtained at fine-scaled propagation distances are needed to closely track the motion of OVs, which is difficult to achieve experimentally. Numerical approach provides a noise-free system and may generate significant interaction characteristics between the OVs at fine-scaled propagation distances. The numerical closed line integral used by Roux and the algorithm based on axial behavior developed by Maallo and Almoro both require scanning before OV can be detected. Hence, we utilized a numerical simulation that is able to detect multiple OVs simulataneously at a fine-scaled propagation distances even as far as 1000 $mm$. The locations of OVs can be determined by getting the intersection of the zeroes of the real and imaginary parts of the complex wave amplitude profile. The validity of numerical results was verified by comparing it to experimental results.

The observation of the dynamical behavior of OVs in HCOB may find potential applications to a more controllable micromanipulation or optical spanner by just adjusting the phase parameters, or a better understanding of reconstruction characteristics of the self-healing property of HCOBs \cite{hermosa2013helico}.

\section{Experiment and Simulations}
\begin{figure}[h!]
\centering
\includegraphics[scale= 0.45]{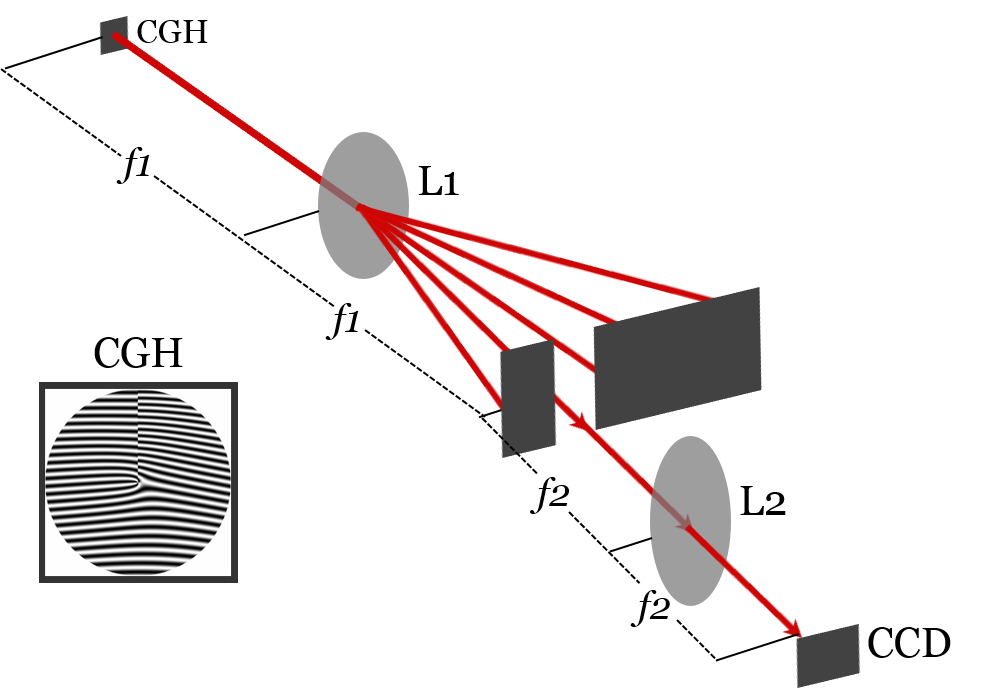}
\caption{\textbf{Experimental setup.} The inset shows a computer-generated hologram (CGH) which was constructed digitally and printed on a mask. The beam splits into different orders after passing through the CGH. Then, the 1$^{st}$ order of diffraction which contains HCOB intensity profile is selected by an aperture. The 2$^{nd}$ lens $L_2$ images the beam onto the CCD camera.}
\label{fig:experimentsetup}
\end{figure}

We present the experimental methods and the simulation steps in this section. Figure \ref{fig:experimentsetup} illustrates the experimental setup. A collimated HeNe laser (wavelength $\lambda$=632.8 $nm$) passes through a computer-generated hologram (CGH) which is placed at the front focal plane of the first lens $L1$ ($f1$ = 500 $mm$). The production of CGH is described elsewhere \cite{chavez2002holographic}.

Intensity of HCOB beam that contains the OVs is acquired experimentally based on the holographic setup used in \cite{chavez2002holographic}. The beam diffracts into different orders after passing through the CGH. HCOB intensity profile can be obtained from the 1$^{st}$ order which is selected by an aperture that is placed at the back focal plane of $L1$. A second lens $L2$ is positioned beyond the aperture at a distance equivalent to its focal length ($f2$ = 250 $mm$). A CCD camera is placed right after the back focal plane of $L2$. The camera captures the near field intensity patterns of the HCOB. We also open the aperture to accommodate the zeroth order so that superposing this with the 1$^{st}$ order will yield interferograms. Interferograms are produced to verify the relative polarities and the magnitudes of OVs. Illustration of the quantification of topological charges by interferograms and confirming the topological charge conservation are presented in \cite{stoyanov2015far}.

Rigorous scalar diffraction theory is utilized for propagating the HCOB numerically. The phase expression given by Equation (\ref{eqn:HCOBphase}) is used as input function of the beam and the wavelength used is 633 $nm$. A complex wave function representing the electric field is acquired after propagation at a set distance. HCOB vortices are located by the intersection between zeroes of real part and imaginary part of the complex electric field. The OVs are located for several propagation distances so that OVs' motions in the transverse spatial structure as the beam propagates can be examined. The propagated functions are obtained from 0.1 $mm$ up to 1000 $mm$ propagation distances with 0.1 $mm$ interval. Lastly, the interferograms can also be generated in simulation by superposing HCOB beams with a plane wave. These interferograms are compared to interferograms that are acquired experimentally in order to examine the agreement between experiment and numerical simulation. Detailed discussion of OV detection technique is presented in the next section.

\section{Detecting optical vortices numerically}
\label{sec:OVdetection}
A 1024 x 1024 pixels frame size is used in the numerical simulation. The side length of this frame is set to 0.8 $cm$. A circular function whose radius is 0.35 $cm$ is multiplied to initial input function to serve as the initial beam size. By propagating the initial function at a certain distance, a complex wave amplitude profile is obtained. 

\begin{figure}[h!]
\centering
\includegraphics[width= 5.0 in]{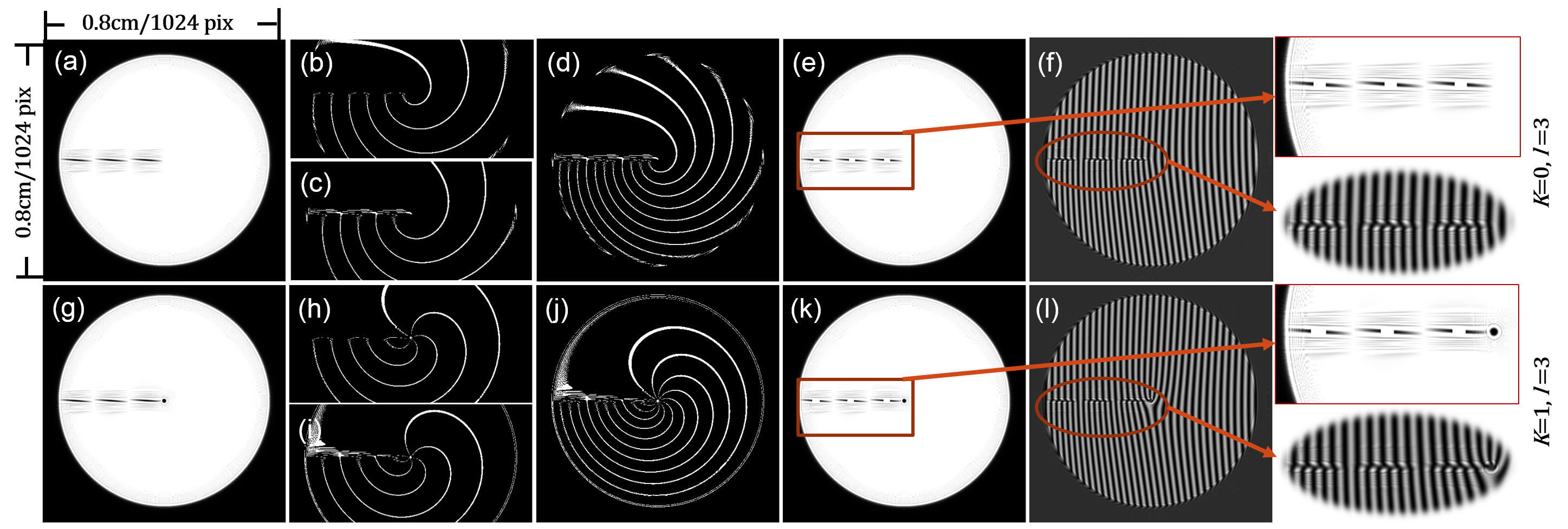}
\caption{OV detection in HCOB beam with $l$=3 and \textit{[1$^{st}$ row images]} $K=0$ or \textit{[2$^{nd}$ row images]} $K=1$ propagated at $z$=1.0 $cm$. The sequence of images from left to right illustrate the ($a$,$g$) intensity profiles, zeroes of ($b$,$h$) \textit{[upper image]} real part and ($c$,$i$) \textit{[lower image]} imaginary part of the complex wave amplitude profile, ($d$,$j$) overlain parts, ($e$,$k$) detected OVs, and ($f$,$l$) interferograms.}
\label{fig:OVdetectz1}
\end{figure}

\begin{figure}[h!]
\centering
\includegraphics[width= 5.0 in]{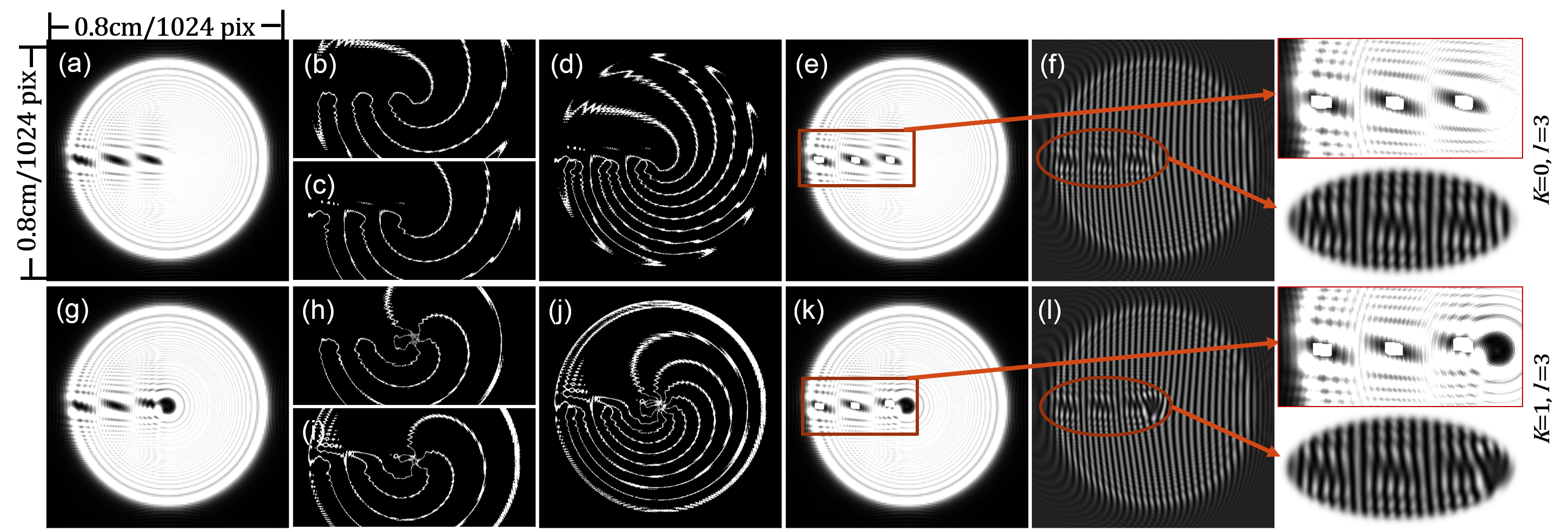}
\caption{OV detection in HCOB beam with $l$=3 and \textit{[1$^{st}$ row images]} $K=0$ or \textit{[2$^{nd}$ row images]} $K=1$ propagated at $z$=20.0 $cm$. The sequence of images from left to right illustrate the ($a$,$g$) intensity profiles, zeroes of ($b$,$h$) \textit{[upper image]} real part and ($c$,$i$) \textit{[lower image]} imaginary part of the complex wave amplitude profile, ($d$,$j$) overlain parts, ($e$,$k$) detected OVs, and ($f$,$l$) interferograms.}
\label{fig:OVdetectz20}
\end{figure}

\begin{figure}[h!]
\centering
\includegraphics[width= 5.0 in]{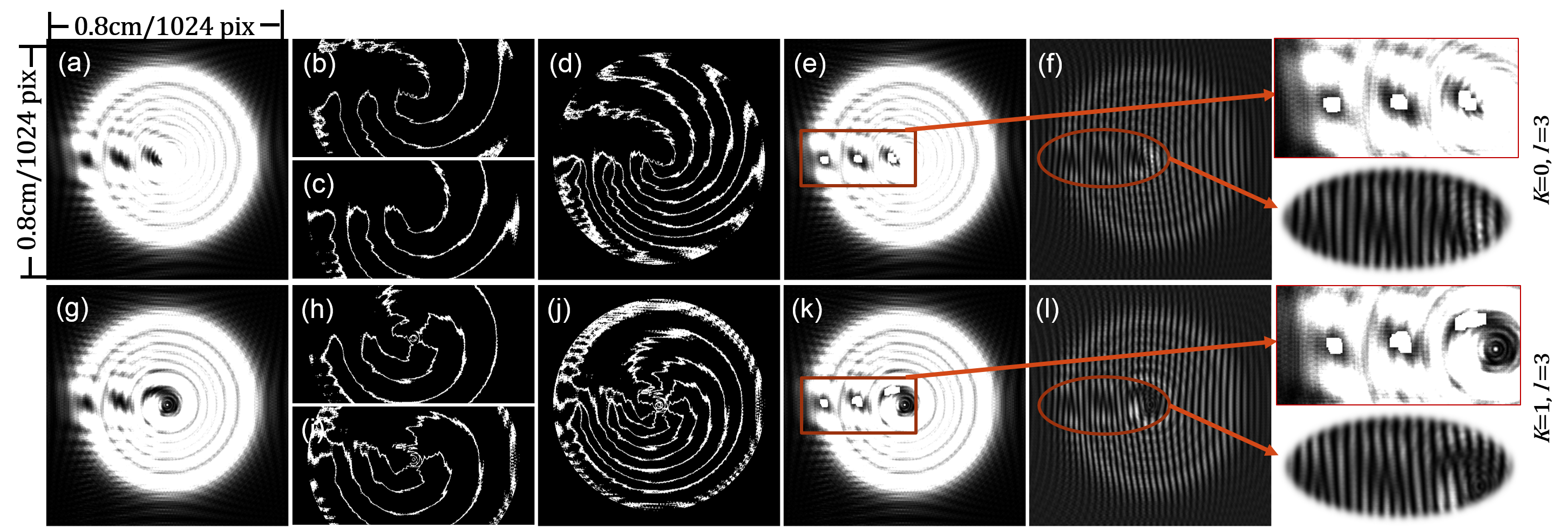}
\caption{OV detection in HCOB beam with $l$=3 and \textit{[1$^{st}$ row images]} $K=0$ or \textit{[2$^{nd}$ row images]} $K=1$ propagated at $z$=80.0 $cm$. The sequence of images from left to right illustrate the ($a$,$g$) intensity profiles, zeroes of ($b$,$h$) \textit{[upper image]} real part and ($c$,$i$) \textit{[lower image]} imaginary part of the complex wave amplitude profile, ($d$,$j$) overlain parts, ($e$,$k$) detected OVs, and ($f$,$l$) interferograms.}
\label{fig:OVdetectz80}
\end{figure}

The OVs are then located by getting the intersection of the zeroes of real and imaginary parts of the complex field. In the numerical simulation, range of values below a chosen threshold pixel value is assigned as zero-valued regions for both the real and imaginary parts. If the threshold value is selected to be very small, we can possibly acquire a single pixel of intersection. This is true for relatively short propagation distances. However, the dark regions on the beam, which correspond to the locations of OVs, expand as the beam propagates. This can be observed in the beams shown in Figures \ref{fig:OVdetectz1}, \ref{fig:OVdetectz20} and \ref{fig:OVdetectz80} which are propagated at 1.0 $cm$, 20.0 $cm$ and 80.0 $cm$ distances, respectively. Thus, for relatively farther propagation distances, enlarged dark regions yield more clustered points of intersection and some inevitably resulted with unconnected points of intersection.

The threshold value for assigning zeroes of real and imaginary parts should be the same for the entire range of propagation distances to achieve uniformity in parameters. We selected the threshold value to be 0.05 which is observed to adequately yield points of intersection. One issue at relatively farther propagation distances is the occurence of unconnected points of intersection that signify the same OV. This is resolved by performing image dilation in which a square with 5-pixeled side is used as structure element to dilate the non-zero valued pixels in the array \cite{haralick1987image}. The image dilation expands a single pixel into the same size of the structure element which enables the connection of unconnected pixels. Since the study aims to observe the dynamical behavior of OVs as far as 1000 $mm$ propagation distance, some resulting unconnected points are much dispersed which require repetition of image dilation. Hence, image dilation is repeated 5 times to ensure the connection of the points and this is done for all the propagation distances for uniformity.

This study requires detection of multiple OVs in a single beam, hence it is helpful to distinguish groups of pixels that signify different OVs. This is done via blob analysis which is a standard image processing when observing dynamics of certain features in an image \cite{carbary2000blob, telagarapu2012novel}. Blob analysis starts with binarized image containing blobs or cluster of connected points. These blobs are assigned with different pixel values for distinction, so that analysis or calculations per blob can be performed. In this work, the regions of interest in the image are the vortices which correspond to dark regions in the beam intensity profile. The blobs in this study are the groups of connected pixel points that signify different OVs. One blob can be isolated in the image by calling its correponding pixel value. The centroid of the blob yields the specific location of an OV. From the coordinates of the centroid, the angular location about the optical axis and radial distance from the optical axis of the OV can be obtained. This calculation is  done for locating all the peripheral OVs at different propagation distances. We first demonstrate the OV detection in Figure \ref{fig:OVdetectz1}. The HCOB phase parameters used are $l$=3 and \textit{[1$^{st}$ row images]} $K=0$ or \textit{[2$^{nd}$ row images]} $K=1$ propagated at $z$=1.0 $cm$.

Three peripheral OVs along the left horizontal discontinuous cut can be noticed for both $K$=0 (\ref{fig:OVdetectz1}$a$) and $K$=1 (\ref{fig:OVdetectz1}$g$). The branching orientations of the zeroes between $K$=0 (\ref{fig:OVdetectz1}$b$ and \ref{fig:OVdetectz1}$c$) and $K$=1 (\ref{fig:OVdetectz1}$h$ and \ref{fig:OVdetectz1}$i$) display hints of the polarities of the OVs. For $K$=0, the branches from the peripheral OVs tend to repel from each other. On the other hand in the presence of a central OV ($K$=1), the branches from peripheral OVs tend to join at the central OV. This indicates that branches of zeroes tend to repel from branch of an OV with same polarity but tend to be attracted to branch of an OV with opposite polarity. The polarities are verified by the fork orientations in the interferograms (\ref{fig:OVdetectz1}$f$ and \ref{fig:OVdetectz1}$l$). Peripheral OVs have same polarity while the central OV is opposite in polarity because of its flipped fork orientation. The overlain zeroes of real and imaginary parts are shown in \ref{fig:OVdetectz1}$d$ and \ref{fig:OVdetectz1}$j$. The intersection from these overlain parts are processed with image dilation and blob analysis. The resulting blobs have detected the OVs as shown in Figures \ref{fig:OVdetectz1}$e$ and \ref{fig:OVdetectz1}$k$. Note that in \ref{fig:OVdetectz1}$k$, the central OV has no corresponding blob since it is masked in the numerical simulation because of the observed location invariance of this OV as the beam propagates. The purpose of removing the blob for central OV is for simplicity during iteration of OV detection at fine-scaled propagation distances. OV detection is also performed at propagation distance of 20.0 $cm$ as illustrated in Figure \ref{fig:OVdetectz20}.

The dark regions on the intensity beam profiles enlarge as the beam propagates from 1.0 $cm$ (\ref{fig:OVdetectz1}$a$, \ref{fig:OVdetectz1}$g$) to 20.0 $cm$ (\ref{fig:OVdetectz20}$a$, \ref{fig:OVdetectz20}$g$). This infers that as the beam propagates it is expected that intersecting zeroes of real and imaginary parts will result to more cluster of points. This is apparent to the blob sizes in \ref{fig:OVdetectz20}$e$ and \ref{fig:OVdetectz20}$k$ after performing image dilation and blob analysis. The 5 repetition of image dilation is useful in this case to connect more distant clustered points yielding larger blobs. The polarities of OVs are consistent to previously observed (same polarities of peripheral OVs that are opposite to the polarity of central OV) from both the branching orientations in the zeroes of complex wave profile shown in \ref{fig:OVdetectz20}$b$, \ref{fig:OVdetectz20}$c$, \ref{fig:OVdetectz20}$h$ and \ref{fig:OVdetectz20}$i$ and the fork orientations in interferograms shown in \ref{fig:OVdetectz20}$f$ and \ref{fig:OVdetectz20}$l$.

The 5 repetition of image dilation is advantageous for beam propagated at farther distance such as this case since the dark regions become larger. The consistency of the branching orientations in the zeroes and the fork orientations in interferograms reveal that the polarities of the OVs do not change as the beam propagates which is expected.

The technique discussed could detect multiple OVs in a single beam even for farther propagation distances. In the data acquisition, we iterated the OV detection in terms of propagation distances within the range of 0.1 $mm$ up to 1000 $mm$ with 0.1 $mm$ interval. This shows that the technique can be used in studying behavioral dynamics of multiples OVs in a single beam at fine-scaled propagation distances.

\section{Results and Discussion}
\label{sec:rnd}
\subsection{Comparison: experimental and numerical results}
\begin{figure}[h!]
\centering
\includegraphics[scale= 0.35]{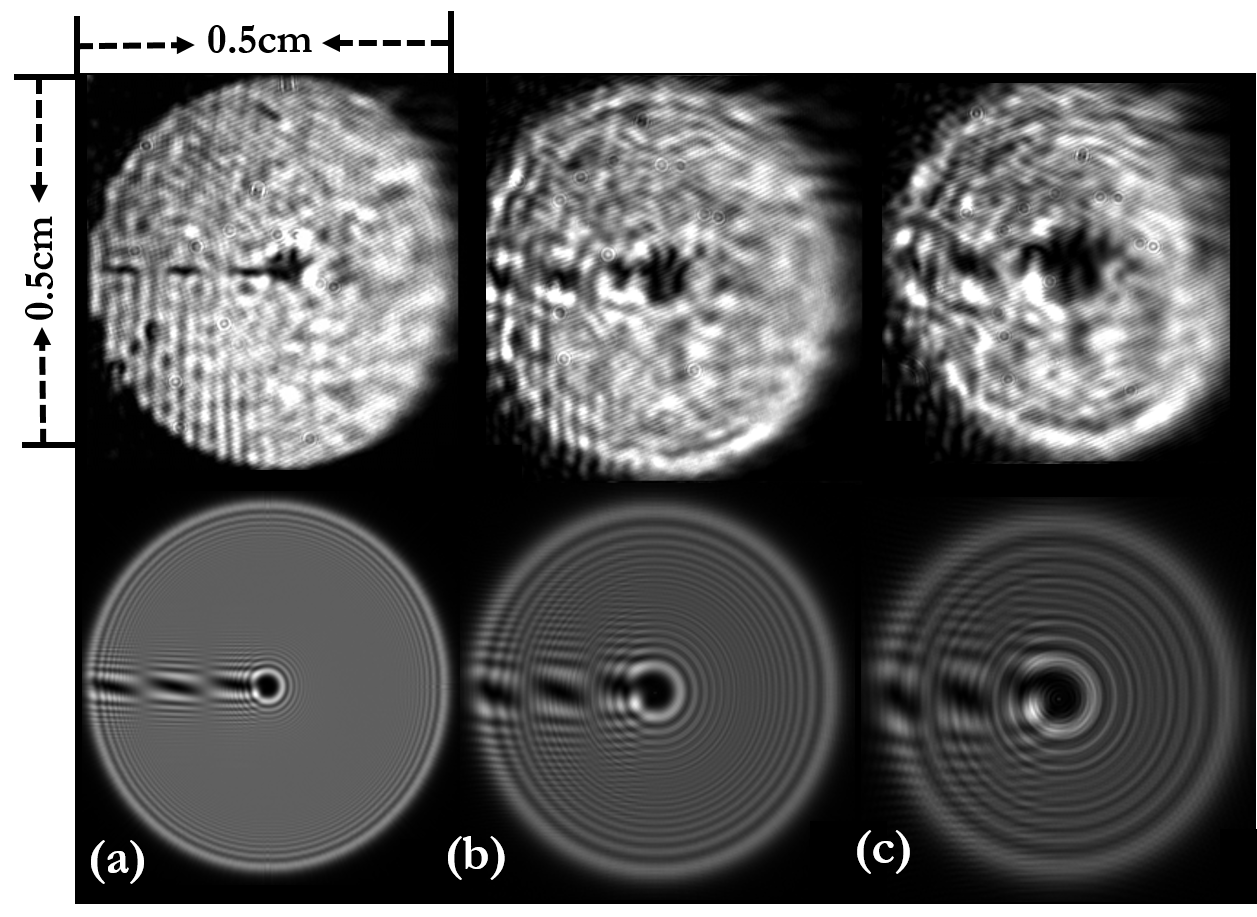}
\caption{Comparison of intensity profiles obtained experimentally (images above) and numerically (images below) for propagation distances (a) near focus $f2$ of 2nd lens, (b) $\sim$13cm from $f2$ and (c) $\sim$27 cm from $f2$.}
\label{fig:wfexpandsim}
\end{figure}

We compare the experimental and numerical beam intensity with phase parameters $l$ and $K$ set to 3 and 1, respectively, in Figure \ref{fig:wfexpandsim}. We display the intensity profiles acquired at several propagation distances (near $f2$, $\sim$13 cm from $f2$, and $\sim$27 cm from $f2$). The near field intensity results between experiment and simulation display similar features. Discontinuous cuts are present along the left horizontal region of the intensity profiles. The cut is noticeably consisting of three peripheral vortices that are aligned to the central vortex. These dark spots enlarge through propagation. The peripheral vortices have eccentricities increasing and major axis slanting as the beam propagates. These evolutions reveal the dynamical behavior of the localized vortices which develop into spiralling intensity at the far field as observed by Alonzo et al. \cite{alonzo2005helico}.

\begin{figure}[h!]
\centering
\includegraphics[scale= 0.35]{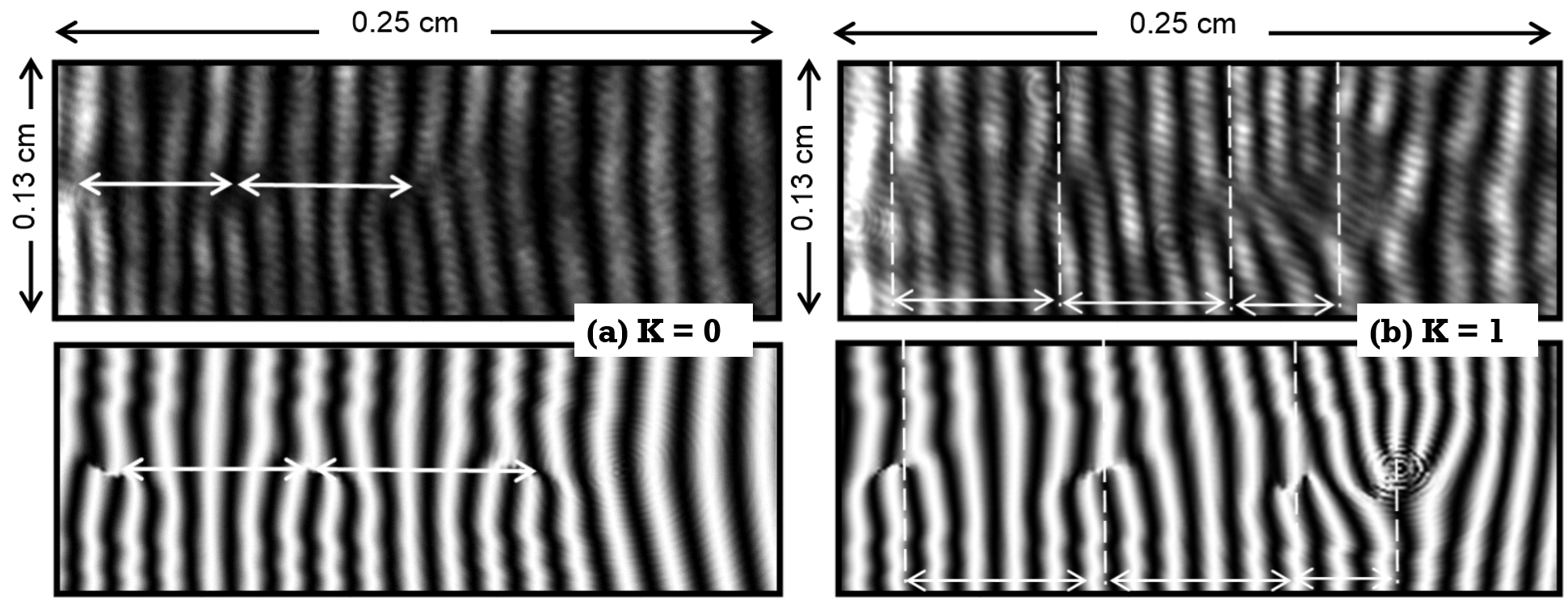}
\caption{HCOB interferograms captured experimentally (images above) and obtained numerically (images below) for $K$=0 and $K$=1.}
\label{fig:expandsim}
\end{figure}

The comparison of experimental and numerical interferograms, for both 0 and 1 values of $K$ is shown in Figure \ref{fig:expandsim}. The central nodes of fork holograms display the locations of the vortices while the arm orientations reveal the polarities. The central nodes that corrrespond to locations of OVs are noticeably aligned peripherally along a single azimuth. The charges of peripheral vortices on both cases are isopolar which is evident by the identical fork orientations. For $K$ equal to 1, the presence of a central vortex whose charge is opposite to the peripheral vortices can be seen. As pertinent to arm counts of fork patterns, the charges of peripheral vortices are all equally 1 while the central vortex is charged 3.

We observed the same correspondence between experimental results and numerical simulation for different values of $l$ ($l$=1,2, 3) and $K$ ($K$=0,1). This indicates that our numerical simulations indeed describe what is being observed. 

\subsection{Simulations: propagation dynamics of OVs}
We illustrate the propagation dynamics of the OVs by location maps. In a location map, the propagation distance axis is projected onto 2D transverse spatial space as indicated by the grayscale values of pixels. Relatively darker pixel dot represents OV location for relatively farther propagation distance. A resulting single pixel dot in the location map indicates that the OV has invariant transverse location or it is not moving along beam propagation.

\begin{figure}[h!]
\centering
\includegraphics[scale= 0.55]{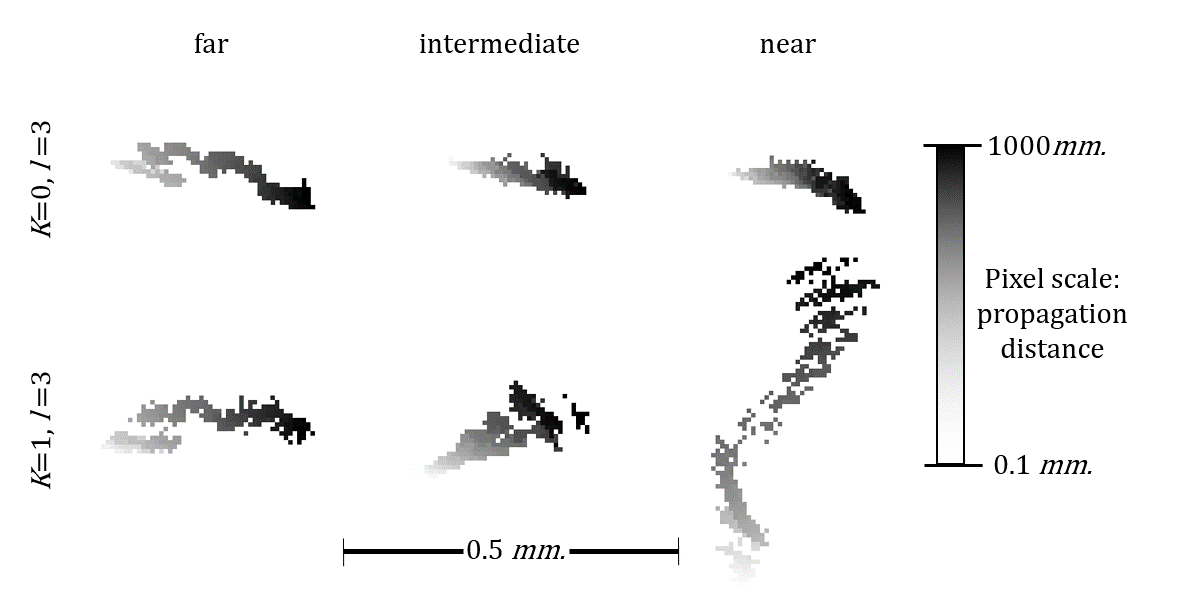}
\caption{\textbf{Location maps}. The propagation axis is projected onto 2D transverse spatial space as the grayscale value of the pixels, with relatively darker pixel as relatively farther propagation distance. In the layout are the location maps of peripheral OVs for $l$=3 and $K$=0,1.}
\label{fig:results1}
\end{figure}

Figure \ref{fig:results1} shows the location maps of three peripheral OVs for both $K$ equal to 0 and 1 with $l$ equal to 3. The farthest propagation distance projected in these maps is 1.0 $m$. These location maps illustrate the dynamics of peripheral OVs in the left horizontal discontinuous cut in Figure \ref{fig:wfexpandsim}. The maps of peripheral OVs are labelled as \textit{near}, \textit{intermediate} and \textit{far} in reference to their radial distances from optical axis. These maps are magnified with the scale shown also in Figure \ref{fig:results1}. No location map is drawn for central OV in the case of $K$=1 since it is observed to be steadily located at the center of the beam. It is evident from the location maps of peripheral OVs that there are dynamical changes in positions of the OVs as the beam propagates.

Two pertinent observations can be extracted from these location maps. First, the vortices are perceived to get attracted toward the optical axis. Second, although not so apparent for farther OVs, the direction of angular displacement is opposite between $K$ equal to 0 and 1. Significant angular displacements can be examined for the case $K$ equal to 1, particularly to the $intermediate$ and  $near$ vortices. The nearer vortices tend to be azimuthally displaced at faster rate. This indicates that the stronger interaction of relatively nearer vortex to the oppositely charged central vortex. The motion of these OVs can be verified mathematically in the same manner as Roux derived the change in location $r(\rho, \theta)$ of an OV with respect to propagation distance $z$. By evaluating the gradient of the general expression of HCOB phase given by Equation (\ref{eqn:HCOBphase}) with an added factor of $-kz$ where $k$ is the wavenumber, the change in location $\vec{r}(\rho,\theta)$ becomes:
\begin{equation}
\frac{d\vec{r}(\rho,\theta)}{dz}=\hat{e_z}+\frac{l\vec{\rho}\times \hat{e_z}}{k\rho^2}(K-\frac{\rho}{\rho_o})-\frac{l\theta\vec{\rho}}{\rho\rho_o}
\label{eqn:drdz}
\end{equation}
where $\hat{e_z}$ and $\vec{\rho}$ denote the unit vector along propagation axis and the separation vector, respectively. Separation vector points to the OV location at $(\rho, \theta)$ coordinate. Note that this expression can describe the interaction of two OVs by setting $\vec{\rho}$ as the difference between vectors defining the locations of two OVs with respect to optical axis. Since no simple mathematical model can describe motion of interacting multiple OVs such as three or more, we can treat Equation (\ref{eqn:drdz}) to approximately describe motion of each OV independently. The existence of the OV as the beam propagates is ensured by the $1^{st}$ term in ($\ref{eqn:drdz}$). The $2^{nd}$ and $3^{rd}$ terms designate the angular and radial behaviors of OV, respectively. The negative sign in the $3^{rd}$ term confirms the inward motion of OVs. The $2^{nd}$ term reveals that some angular displacements of peripheral OVs can be observed with ($K$=1) or without ($K$=0) the central OV. Moreover, the direction of angular displacement is opposite between beams with and without the central OV. These angular behavior of OVs are observed in the location maps.

\begin{figure}[h!]
\centering
\includegraphics[scale= 0.3]{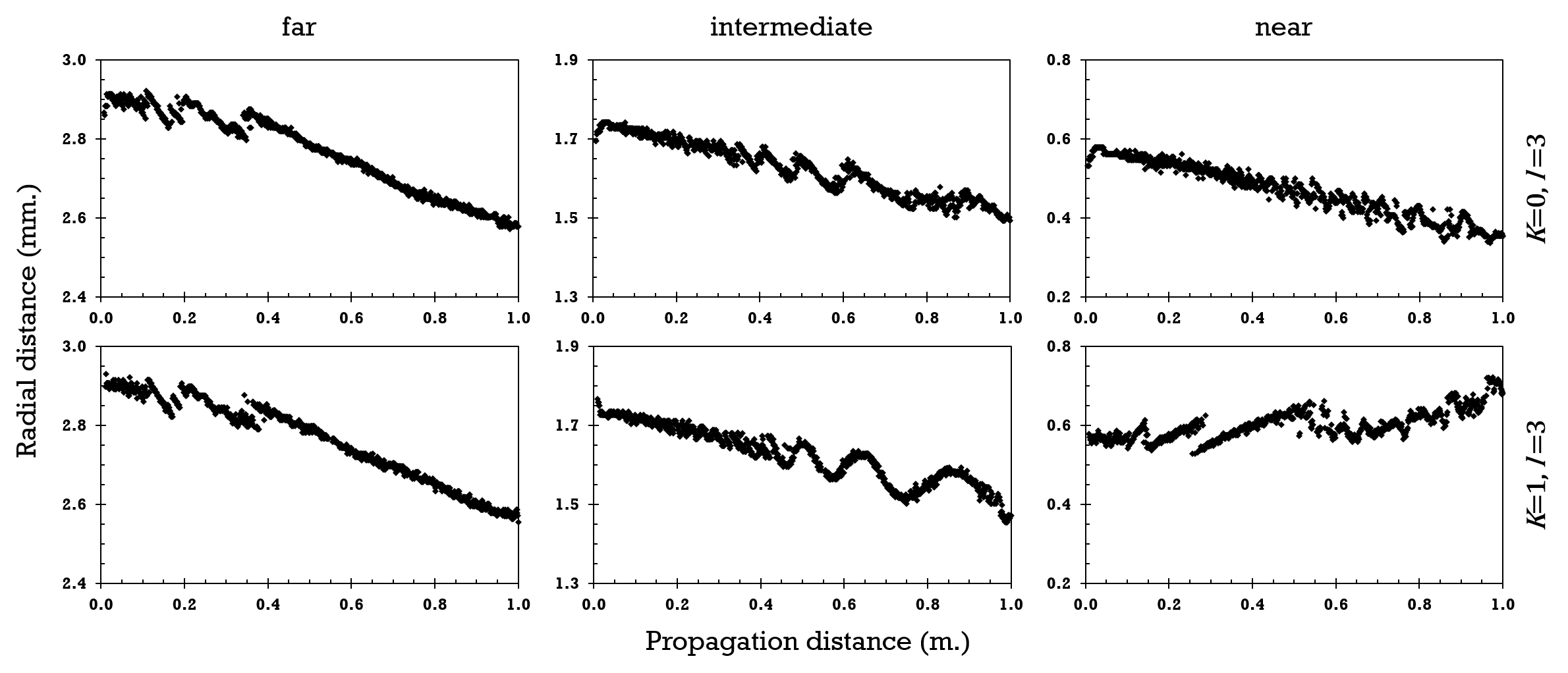}
\caption{\textbf{Radial motion}. Plots of radial distance versus propagation distance of peripheral OVs for $l$=3 and $K$=0,1.}
\label{fig:radial}
\end{figure}

\begin{figure}[h!]
\centering
\includegraphics[scale= 0.3]{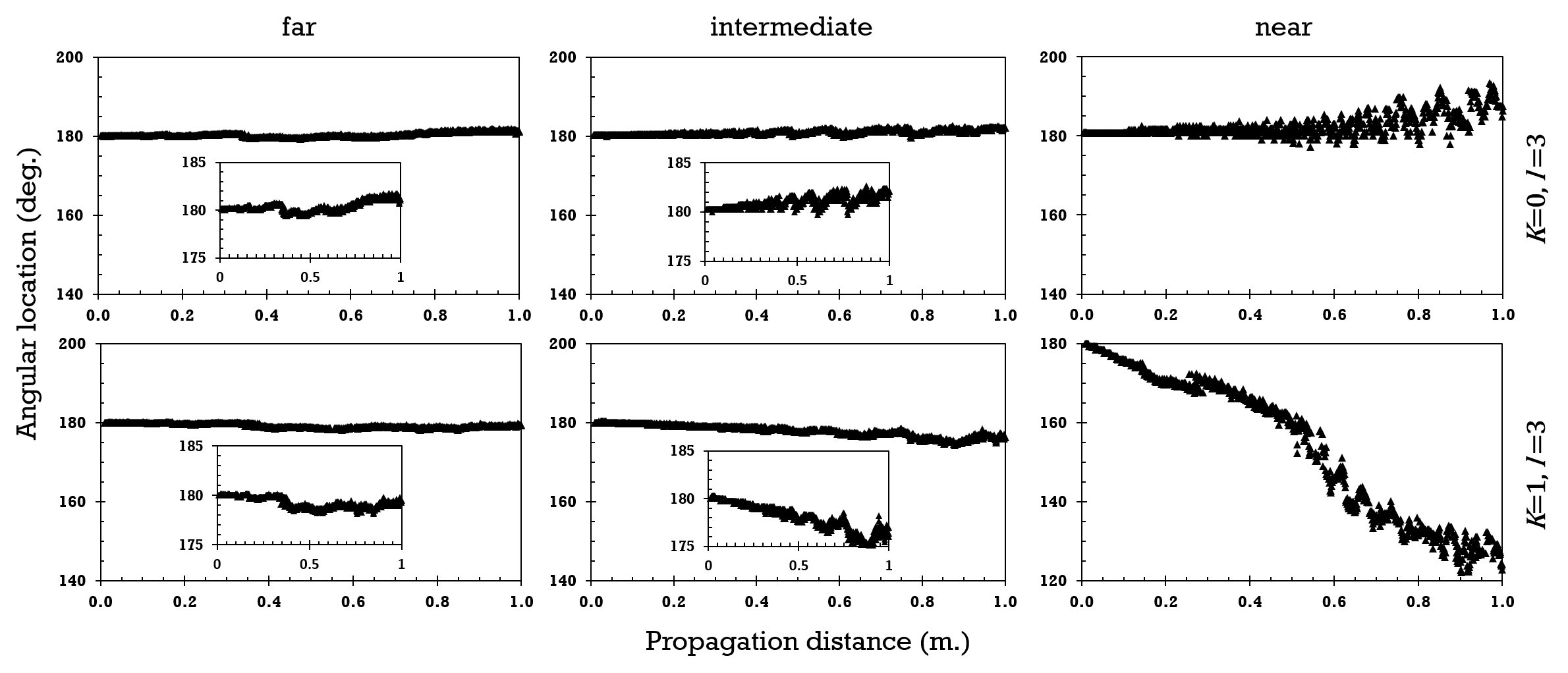}
\caption{\textbf{Angular motion}. Plots of angular location versus propagation distance of peripheral OVs for $l$=3 and $K$=0,1.}
\label{fig:angular}
\end{figure}

The location maps are helpful for visualizing motion of OVs in the transverse spatial space as the beam propagates. To closely examine with quantitative analysis, the location maps are decomposed to plots comprising the radial and angular displacement versus the propagation distance as presented in Figures \ref{fig:radial} and \ref{fig:angular}, respectively.

For the case of $l$=3 and $K$=0, all 3 peripheral OVs are found to have inward motion as can be observed from upper plots in Figure \ref{fig:radial} wherein the net radial distances decrease after 1.0 $m$ propagation. This result agrees with the radial change of location with respect to propagation distance given by Equation (\ref{eqn:drdz}). Similarly for $l$=3 and $K$=1, peripheral OVs have inward motion except for the $near$ OV as shown by lower plots in Figure \ref{fig:radial}. Both results indicate that peripheral OVs are somewhat attracted towards the optical axis except for the $near$ OV in $K$=1. The radial plots between $K$ equal to 0 and 1 display almost the same trends for both $intermediate$ and $far$ OVs. On the other hand, the trends in radial plots of the $near$ OV differ between $K$ equal to 0 and 1, in which dynamics of $near$ OV is influenced by the presence of a central OV. These observations infer that the presence of strongly charged central OV greatly influences the dynamics of nearest OV neighbor and negligibly affects the dynamics of the farthest OV neighbor. Lastly, we remark on the radial behavior when undulation is observed in the $intermediate$ OV for both $K$ equal to 0 and 1.  This can be interpreted as the push-pull induced motion to an OV ($intermediate$) due to the OVs ($near$ and $far$) that surround it. This undulation cannot be modelled by Equation (\ref{eqn:drdz}) and an interaction term between the OVs may be warranted. The dynamics of the OVs, especially the $near$ OV, are further investigated for angular displacaments as the beam propagates. 

In the absence of a central OV ($K$=0), the peripheral OVs tend to have positive azimuthal displacements as apparent in the trends of the upper plots in Figure \ref{fig:angular}. However, the angular displacements for $intermediate$ and $far$ OVs are minute compared to pertinent net angular displacement for $near$ OV after 1.0 $m$ propagation. The higher degree of net angular displacement of $near$ OV compared to $intermediate$ and $far$ OVs are also observed in the case with the presence of a central OV ($K$=1) as can be seen in the lower plots in Figure \ref{fig:angular}. It can also be noticed that the directions of angular displacements of OVs between $K$ equal to 0 and 1 are opposite. The observed reverse direction is expected as the sign in 2$^{nd}$ term in the expression (\ref{eqn:drdz}) switches when we set $K$ to either 0 or 1. The results manifest that $near$ OV, either with or without a central OV, tend to be angularly displaced at higher magnitude compared to farther OVs. The central OV is strongly interacting with the nearest OV neighbor. The strongly charged central OV prevents the inward motion but heightens the gyration of $near$ OV. 


We also investigated the dynamics of peripheral OVs for higher $l$ values such as 4, 5 and 6, in which similar results compared to $l$ equal to 3 are observed. Inward motion toward the optical axis is seen for peripheral OVs for either $K$ equal to 0 or 1. The directions of angular displacements of peripheral OVs are opposite between HCOBs with central OV and without central OV. It is also observed for HCOBs with higher $l$ values that the presence of central OV triggers the inward motion and induces significant angular displacement to the nearest peripheral OV. We inspected in $l$ equal to 4 that the undulation also occured in the intermediate peripheral OVs but with lower amplitude and higher frequency. However, as $l$ value goes higher such as 5 and 6, this undulation becomes less apparent. Moreover for higher $l$ values, the farther the OV from the optical axis the less it is affected by the presence of a central OV since it is observed that direction of angular displacement of this OV is same with the case for no central OV. The dynamics of more peripheral OVs is beyond the scope of this study, although it will be interesting to investigate and detail the behavior.

\section{Summary and conclusion}
Peripheral OVs of HCOB as the beam propagates are observed to have dynamical behavior in the transverse spatial space. The dynamics is examined by the plots of radial distance and angular displacement versus the propagation distance. Inward motion toward the optical axis is observed for the peripheral OVs as the beam propagates for both $K$ equal to 0 and 1. The direction of angular displacement is opposite between beams with and without a central OV. The nearest OV to optical axis has relatively higher net angular displacements compared to farther OVs as observed for both the case $K$ equal to 0 and 1. Peculiar result is investigated for $near$ OV in case of $K$=1 because it strongly interacts with the central OV. Different from the rest peripheral OV, the $near$ OV is found to have heightened angular motion about the strongly charged central OV. The gyration of peripheral OVs with incurred inward motion as the beam propagates explain the spiral intensity distribution of HCOB at the far field. This propagation dynamics can be helpful for a more controllable rotational positioning of assymetrical particles.

\section*{Acknowledgement}
N. Hermosa received a relocation grant as a Balik PhD from the Office of the Vice President for Academic Affairs, UP system.

\section*{References}

\bibliography{mybibfile}

\begin{thebibliography}{10}
\expandafter\ifx\csname url\endcsname\relax
  \def\url#1{\texttt{#1}}\fi
\expandafter\ifx\csname urlprefix\endcsname\relax\def\urlprefix{URL }\fi
\expandafter\ifx\csname href\endcsname\relax
  \def\href#1#2{#2} \def\path#1{#1}\fi

\bibitem{nye1974dislocations}
J.~Nye, M.~Berry, Dislocations in wave trains, in: Proceedings of the Royal
  Society of London A: Mathematical, Physical and Engineering Sciences, Vol.
  336, The Royal Society, 1974, pp. 165--190.

\bibitem{allen1992orbital}
L.~Allen, M.~W. Beijersbergen, R.~Spreeuw, J.~Woerdman, Orbital angular
  momentum of light and the transformation of laguerre-gaussian laser modes,
  Physical Review A 45~(11) (1992) 8185.

\bibitem{volke2002orbital}
K.~Volke-Sepulveda, V.~Garc{\'e}s-Ch{\'a}vez, S.~Ch{\'a}vez-Cerda, J.~Arlt,
  K.~Dholakia, Orbital angular momentum of a high-order bessel light beam,
  Journal of Optics B: Quantum and Semiclassical Optics 4~(2) (2002) S82.

\bibitem{tao2004experimental}
S.~H. Tao, W.~M. Lee, X.~Yuan, Experimental study of holographic generation of
  fractional bessel beams, Applied optics 43~(1) (2004) 122--126.

\bibitem{gotte2008light}
J.~B. G{\"o}tte, K.~O’Holleran, D.~Preece, F.~Flossmann, S.~Franke-Arnold,
  S.~M. Barnett, M.~J. Padgett, Light beams with fractional orbital angular
  momentum and their vortex structure, Optics express 16~(2) (2008) 993--1006.

\bibitem{gibson2004free}
G.~Gibson, J.~Courtial, M.~Padgett, M.~Vasnetsov, V.~Pas'~ko, S.~Barnett,
  S.~Franke-Arnold, Free-space information transfer using light beams carrying
  orbital angular momentum, Optics Express 12~(22) (2004) 5448--5456.

\bibitem{cojoc2005laser}
D.~Cojoc, V.~Garbin, E.~Ferrari, L.~Businaro, F.~Romanato, E.~Di~Fabrizio,
  Laser trapping and micro-manipulation using optical vortices, Microelectronic
  Engineering 78 (2005) 125--131.

\bibitem{lee2004optical}
W.~Lee, X.-C. Yuan, W.~Cheong, Optical vortex beam shaping by use of highly
  efficient irregular spiral phase plates for optical micromanipulation, Optics
  letters 29~(15) (2004) 1796--1798.

\bibitem{roux1995dynamical}
F.~S. Roux, Dynamical behavior of optical vortices, JOSA B 12~(7) (1995)
  1215--1221.

\bibitem{vaupel1996hydrodynamic}
M.~Vaupel, K.~Staliunas, C.~Weiss, Hydrodynamic phenomena in laser physics:
  Modes with flow and vortices behind an obstacle in an optical channel,
  Physical Review A 54~(1) (1996) 880.

\bibitem{rozas1997experimental}
D.~Rozas, Z.~Sacks, G.~Swartzlander~Jr, Experimental observation of fluidlike
  motion of optical vortices, Physical review letters 79~(18) (1997) 3399.

\bibitem{ahmad2002statistical}
F.~Ahmad, W.~C. Saslaw, N.~I. Bhat, Statistical mechanics of the cosmological
  many-body problem, The Astrophysical Journal 571~(2) (2002) 576.

\bibitem{sah1958electron}
C.-T. Sah, W.~Shockley, Electron-hole recombination statistics in
  semiconductors through flaws with many charge conditions, Physical Review
  109~(4) (1958) 1103.

\bibitem{hermosa2007phase}
N.~P. Hermosa, C.~O. Manaois, Phase structure of helico-conical optical beams,
  Optics communications 271~(1) (2007) 178--183.

\bibitem{singh2013vortices}
B.~K. Singh, D.~Mehta, P.~Senthilkumaran, Vortices in helico-conical beam and
  fractional vortex beam, in: Recent Advances in Photonics (WRAP), 2013
  Workshop on, IEEE, 2013, pp. 1--2.

\bibitem{maallo2011numerical}
A.~M.~S. Maallo, P.~F. Almoro, et~al., Numerical correction of optical vortex
  using a wrapped phase map analysis algorithm, Optics letters 36~(7) (2011)
  1251--1253.

\bibitem{murphy2010experimental}
M.~Kevin, D.~Burke, N.~Devaney, C.~Dainty, Experimental detection of optical
  vortices with a shack-hartmann wavefront sensor, Optics Express 18~(15).

\bibitem{ricci2012instability}
F.~Ricci, W.~L{\''o}ffler, M.~van Exter, Instability of higher-order optical
  vortices analyzed with a multi-pinhole interferometer, Optics express 20~(20)
  (2012) 22961--22975.

\bibitem{hermosa2013helico}
N.~Hermosa, C.~Rosales-Guzm{\'a}n, J.~Torres, Helico-conical optical beams
  self-heal, Optics letters 38~(3) (2013) 383--385.

\bibitem{chavez2002holographic}
S.~Ch{\'a}vez-Cerda, M.~Padgett, I.~Allison, G.~New, J.~C. Guti{\'e}rrez-Vega,
  A.~O’Neil, I.~MacVicar, J.~Courtial, Holographic generation and orbital
  angular momentum of high-order mathieu beams, Journal of Optics B: Quantum
  and Semiclassical Optics 4~(2) (2002) S52.

\bibitem{stoyanov2015far}
L.~Stoyanov, S.~Topuzoski, I.~Stefanov, L.~Janicijevic, A.~Dreischuh, Far field
  diffraction of an optical vortex beam by a fork-shaped grating, Optics
  Communications 350 (2015) 301--308.

\bibitem{haralick1987image}
R.~M. Haralick, S.~R. Sternberg, X.~Zhuang, Image analysis using mathematical
  morphology, Pattern Analysis and Machine Intelligence, IEEE Transactions
  on~(4) (1987) 532--550.

\bibitem{carbary2000blob}
J.~Carbary, K.~Liou, A.~Lui, P.~Newell, C.~Meng, “blob” analysis of auroral
  substorm dynamics, Journal of Geophysical Research: Space Physics
  (1978--2012) 105~(A7) (2000) 16083--16091.

\bibitem{telagarapu2012novel}
P.~Telagarapu, M.~N. Rao, G.~Suresh, A novel traffic-tracking system using
  morphological and blob analysis, in: Computing, Communication and
  Applications (ICCCA), 2012 International Conference on, IEEE, 2012, pp. 1--4.

\bibitem{alonzo2005helico}
C.~Alonzo, P.~J. Rodrigo, J.~Gl{\"u}ckstad, Helico-conical optical beams: a
  product of helical and conical phase fronts, Optics express 13~(5) (2005)
  1749--1760.

\end{thebibliography}

\end{document}